\documentclass{amsproc}
\usepackage{graphicx}
\usepackage{hyperref}
\usepackage{appendix}

\usepackage{amscd}
\usepackage{amsmath}
\usepackage{epsfig}
\usepackage{amsfonts}
\usepackage{amssymb}
\usepackage{hyperref}

\usepackage{pdfpages}
\usepackage[thinc]{esdiff}
\usepackage{accents}
\usepackage{nicefrac}

\theoremstyle{definition}

\theoremstyle{remark}

\numberwithin{equation}{section}

\allowdisplaybreaks
\input{tcilatex}

\newcommand{\iun}{\mathrm{i}\mkern1mu}

\begin{document}
\title[On Fermi's Oversight]
{On a Seldom Oversight in Fermi's Calculations: \\
Seventy Years Later\/}
\author{Sergei K. Suslov}
\address{School of Mathematical and Statistical Sciences, Arizona State
University, P.~O.\ Box 871804, Tempe, AZ 85287-1804, U.S.A.}
\email{sergei@asu.edu}
\date{today}

\begin{abstract}
We discuss an unfortunate mistake, for a Dirac free particle, in the last Fermi lecture notes on quantum mechanics,
in a course given at the University of Chicago in winter and spring of 1954.
As is demonstrated, the correct result can be obtained by a simple matrix multiplication.
An attempt to collect a relevant bibliography is made.

\noindent
%
%
\end{abstract}

\maketitle

\medskip

{\scriptsize{I studied mathematics with passion because I considered it necessary for the study of physics, \it{to which I want to dedicate myself exclusively\/}.}}
\begin{flushright}
\scriptsize{{Comment on his mathematical education by young Enrico~Fermi, Nobel Prize in Physics 1938\/} \cite{Segre}}
\end{flushright}

\smallskip

{\it{ --- To Dr.~Miranda Materi\/}}

\smallskip

An elementary introduction to the Dirac equation and its free particle solutions, based on the notes of the great Master, can serve as a valuable resource in the classroom for mathematicians, physicists, and engineers. Our approach relies solely on the basic algebra of $4 \times 4\/$ partitioned matrices. All results can be readily verified either manually or with the aid of a computer algebra system.

\section{Enrico Fermi as a Great Mentor and Scientist\/}

Several of Fermi's biographies \cite{Bruzz, Schwartz, Segre, SegreHoerlin},
recollections \cite{FermiLaura, PontecorvoFermi, PontecorvoFermi93, PontecorvoFermi04, Segre88},
historical investigations \cite{Acocellaetal04, Moreira22, GuerraRobotti06, GuerraRobotti12, GuerraRobotti20, GuerraRobotti09, GuerraRobotti, LeoneRobSeg00},
the collected works in two volumes \cite{FermiItaly, FermiUSA}, as well as
research \cite{AndersonAllison} and lecture notes \cite{FermiNuclRoma, Fermi},
were published after his untimely death in November of 1954.
Early in his distinguished career, Fermi held a temporary job at the University of Rome, where he taught a mathematics course for chemists and biologists.
From 1924 to 1926, he lectured on mathematical physics and mechanics at the University of Florence.
During this period, Fermi studied Schr{\"{o}}dinger's theory in depth through the original publications
and privately introduced it to his students in seminars.
Later, he adapted some of Dirac's papers into a more accessible format, partly for didactic purposes \cite{Fermi, Segre}.
Subsequently, Fermi was appointed as a professor of theoretical physics at the University of Rome,
the first chair of this kind in Italy, where he taught for 12 years, starting in 1926
\cite{Bruzz, Schwartz, Segre, Segre88}.
%

%
During this period, Fermi established a solid foundation for education in modern physics both in Italy and internationally. He delivered numerous popular lectures and seminars, authored a textbook, and contributed articles to the \textit{Enciclopedia Italiana Treccani} \cite{FermiItaly}.
His teaching style and charismatic personality drew many talented students to the physics department.
An entire generation of physicists worldwide studied the quantum theory of radiation through his seminal review article \cite{FermiRad}, which was based on lectures delivered during the summer of 1930 at the University of Michigan, Ann Arbor, when Fermi visited the United States for the first time.
Students recalled the remarkable atmosphere of immense enthusiasm and unwavering dedication to physics, where they forged lifelong friendships.
As early as 1928, Fermi made little use of books; a collection of mathematical formulas and tables of physical constants were nearly the only reference materials in his office. If a complicated equation was needed for his research or teaching, Fermi could derive it himself -- often faster than his students could locate the result in library books \cite{Schwartz, Segre, Segre88}.
At that time, his name became associated with the Thomas-Fermi model of the atom and the Fermi-Dirac statistics. 
%

%
International recognition of Fermi's work on quantum statistics in Italy followed the Volta Congress on Lake Como in September 1927 \cite{SegreHoerlin}.
Enrico Fermi was the first Italian to present a talk at the prestigious Sixth Solvay Conference on Magnetism in 1930, owing to his work on hyperfine structures \cite{FermiSol, GuerraRobotti}.
In the fall of 1931, Fermi served as General Secretary of an International Conference on Nuclear Physics, held in Rome from October 11 to October 18 -- the first international conference in this field.
A year later, he delivered a report on the current state of nuclear physics at the Paris Conference, representing Italy \cite{GuerraRobotti}. (The neutron and positron were also discovered in 1932.)
%

At the end of 1933, Fermi authored his renowned article on the explanation of beta decay.
He submitted a letter to \textit{Nature} outlining his theory, but it was rejected%
\footnote{The editors felt that ``it contained abstract speculations too remote from physical reality to be of interest
to the reader \cite{SegreHoerlin}''\/.}%
. Instead, the extended article \emph{Tentative theory
of beta rays\/} was published in \textit{Nuovo Cimento} in Italian and in
\textit{{Zeitschrift f\"{u}r Physik\/}} in German \cite{FermiItaly, FermiLaura, Segre}.
The groundbreaking neutron bombardment experiments, systematically documented in \textit{Ricerca Scientifica} letters during the summer of 1934, were described as an ``... escape from the sphere of theoretical physics'' (in Rutherford's own words \cite{Segre}
{\footnote{see \cite{GuerraRobotti}, page~229, for the original letter.}}).
The neutron work completed by that summer was summarized in an article communicated by Lord Rutherford to the Royal Society of London.
The discovery of slow neutron effects later in 1934 remains a remarkable milestone in the history of nuclear physics \cite{Bruzz, FermiNuclRoma, FermiUSA, FermiLaura, GuerraRobotti20, GuerraRobotti09, GuerraRobotti,
PontecorvoFermi, Schwartz, Segre, SegreHoerlin}.
%

Fermi's visit to Ann Arbor was a great scientific success, and he returned there during the summers of 1933 and 1935. In 1934, he embarked on a lecture tour through Argentina, Brazil, and Uruguay \cite{Moreira22, Schwartz}.
In 1936, Fermi visited Columbia University for the summer session. The following year, he spent the summer in California \cite{Schwartz}, lecturing at Berkeley and Stanford, and  driving back to New York across the entire country.
Through these visits, he developed an appreciation for America and the new opportunities it offered.
Fermi ultimately relocated to the United States soon after receiving the Nobel Prize in Physics on December 10, 1938.
His major contributions to the Manhattan Project \cite{ReedMP-4} and work with the Atomic Energy Commission \cite{ABuck} are well-documented \cite{Katz, Schwartz, Segre, SegreHoerlin}.
%

%
Throughout his life, Fermi maintained a strong passion for teaching. He conducted numerous courses and seminars at the University of Rome, University of Michigan, Stanford University, Columbia University, Los Alamos Lab, and the University of Chicago.
In the winter and spring semesters of his final year, before his untimely death in November, Fermi taught his last introductory quantum mechanics course at the University of Chicago \cite{Fermi}. 
During the summer of 1954, he traveled to Europe for the final time. 
While there, Fermi presented a course on pions and nucleons at the Villa Monastero in Varenna on Lake Como. 
This course was part of the summer school organized by the Italian Physical Society, which now bears his name. 
Before that Enrico attended the French summer school at Les Houches, near Chamonix, where he delivered additional lectures \cite{Schwartz, Segre, SegreHoerlin}.%
\footnote{%
From Britannica: Enrico Fermi, (born Sept. 29, 1901, Rome, Italy -- died
Nov. 28, 1954, Chicago, Illinois, U.~S.~A.), Italian-born American scientist
who was one of the chief architects of the nuclear age. He developed the
mathematical statistics required to clarify a large class of subatomic
phenomena, explored nuclear transformations caused by neutrons, and directed
the first controlled chain reaction involving nuclear fission. He was
awarded the 1938 Nobel Prize for Physics, and the Enrico Fermi Award of the
U.S. Department of Energy is given in his honor. Fermilab, the National
Accelerator Laboratory, in Illinois, is named for him, as is fermium,
element number 100. {\url{https://www.britannica.com/biography/Enrico-Fermi}}%
}

\begin{figure}[h!]
	\includegraphics[width=0.65\linewidth]{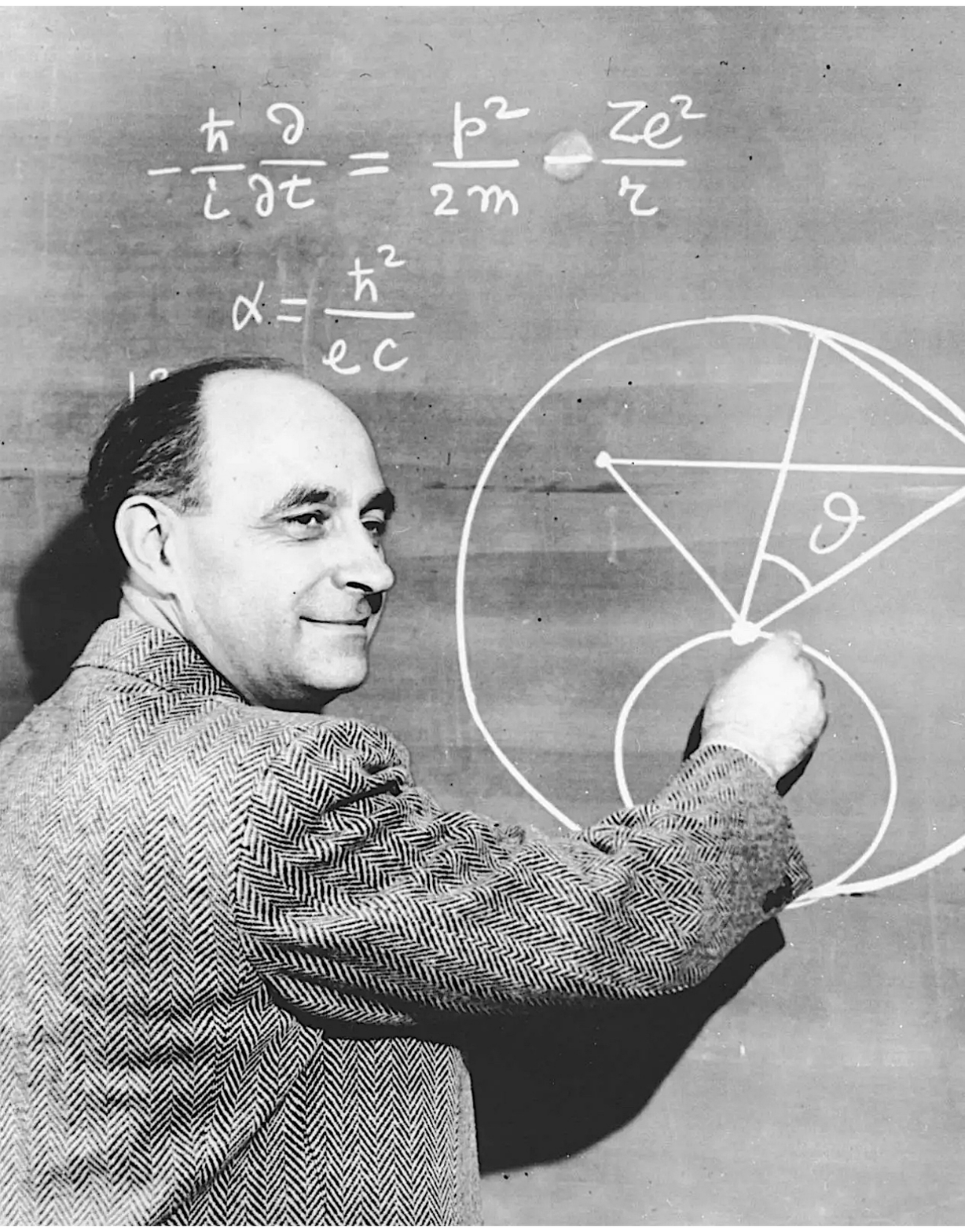}
	\caption{Enrico Fermi teaching quantum mechanics. This photo and its counterpart are taken at the University of Chicago 
     on 26~March 1948 \cite{Huber} .
	{\it{Courtesy of Argonne National Lab\/.}} (The second formula 
    is, most likely, his idea of a joke \cite{Huber, Lien, Schwartz}).)
    \url{https://science.osti.gov/fermi/The-Life-of-Enrico-Fermi/formula}%
}
\end{figure}
\begin{figure}[h!]
	\includegraphics[width=0.7915\linewidth]{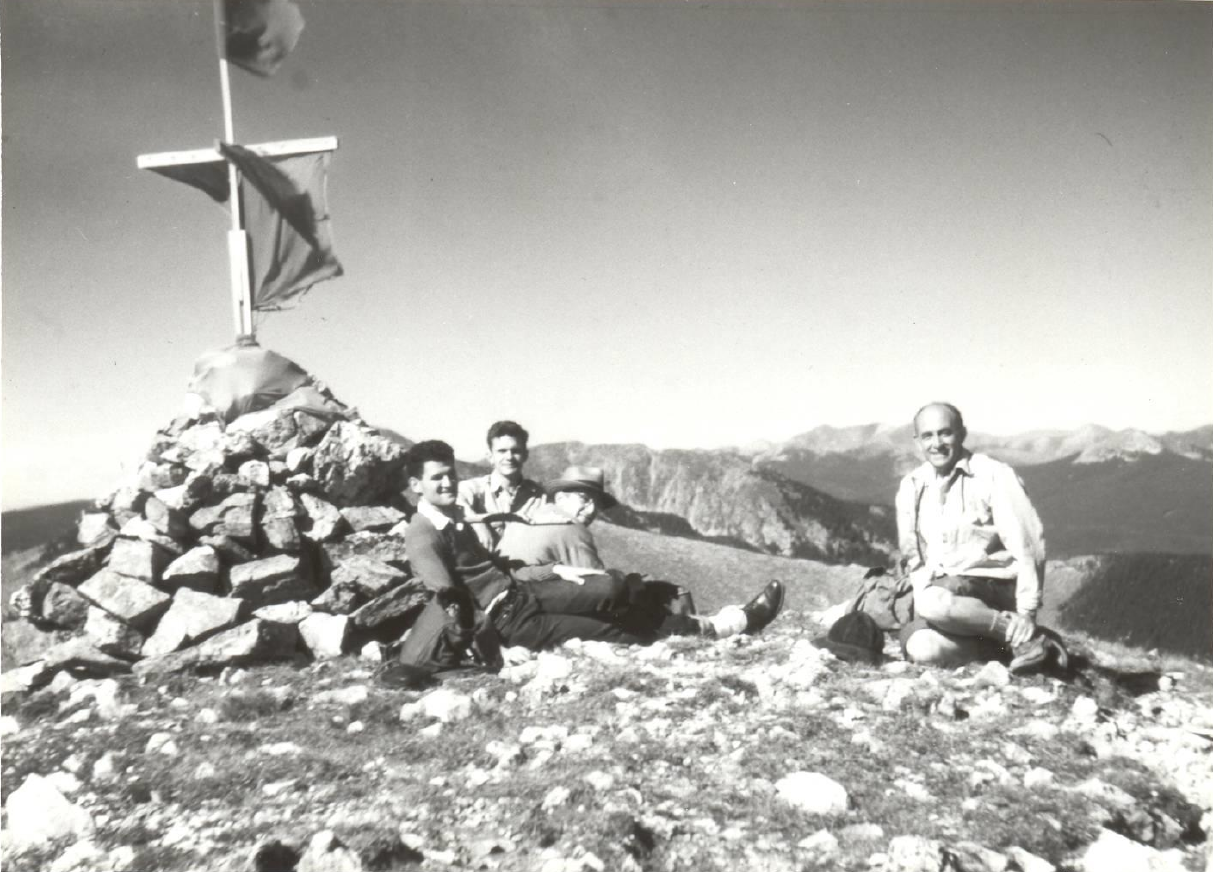}
	\caption{Enrico Fermi hiking with students in New Mexico.
	{\it{Courtesy of Peter Lax\/.}}}
\end{figure}

\section{Dirac's Equation for a Free Particle and Its Simplest Matrix Solutions\/}

The magnificent relativistic wave equation formulated by Dirac for a free spin-$1/2$ particle, 
with the electron as an example,
is given by \cite{AkhBer, BerLifPit, Fermi, Moskalev, PeskSchroe, Schiff, Susetal20}:
\begin{equation}
	\iun \hbar \diffp{}{t} \psi(\mathbf{r}, t)
	=
	\widehat{H} \psi(\mathbf{r}, t),
	\label{d1}
\end{equation}
where $\iun$ is the imaginary unit, $\hslash$ is the reduced Planck constant, $\mathbf{r}$ is a three-component vector representing position in space, and $t$ is time. 
Here, $\widehat{H}$ is a relativistic Hamiltonian operator and $\psi$ is a complex-valued function to be discussed below.

Let us recall the standard Pauli matrices, given by
\begin{equation}
	{\sigma}_1 =
	\begin{pmatrix}
		0 & 1 \\
		1 & 0
	\end{pmatrix}\/ ,
	\qquad
	{\sigma}_2 =
	\begin{pmatrix}
		0 & -\iun \\
		\iun & 0
	\end{pmatrix}\/ ,
	\qquad
	{\sigma}_3 =
	\begin{pmatrix}
		1 & 0 \\
		0 & -1
	\end{pmatrix}\/ ,
	\label{d4}
\end{equation}
and denote
\begin{equation}
	O := 
	\begin{pmatrix}
		0 & 0 \\
		0 & 0
	\end{pmatrix} ,
	\qquad
	I= 
    {I}_2 :=
	\begin{pmatrix}
		1 & 0 \\
		0 & 1
	\end{pmatrix} \/ .
	\label{d01}
\end{equation}
Using these $2 \times 2$ matrices as building blocks, one can construct
the following $4 \times 4$ (Hermitian) Dirac matrices:
\begin{equation}
	\boldsymbol{{\alpha}}
	=
	\begin{pmatrix}
		O 
        & \boldsymbol{{\sigma}} \\
		\boldsymbol{{\sigma}} & O 
	\end{pmatrix} , \qquad
{\beta} 	=
	\begin{pmatrix}
		I 
        & O \\ 
        O & -I
	\end{pmatrix} .
	\label{d3}
\end{equation}
Finally, the standard Hamiltonian of the above evolutionary Schr{\"o}dinger-type equation takes the form
\begin{equation}
	\widehat{H}
	=
	c
	\left(
		\boldsymbol{\alpha}
		\cdot
		\widehat{\boldsymbol{p}}
	\right)
	+mc^{2} {\beta}\/ ,
	\label{d2}
\end{equation}
where $\boldsymbol{\alpha} \cdot \widehat{\boldsymbol{p}} =
{\alpha_1} \widehat{p}_1 + {\alpha_2}
\widehat{p}_2 + {\alpha _{3}} \widehat{p}_{3}$ with the
linear momentum operator $\widehat{\boldsymbol{p}}=-\iun\hbar
\nabla$. 
As usual,
$m$ is the rest mass and $c$ is the speed of light in vacuum.

The relativistic electron has a four-component (bi-spinor) complex-valued wave function
\begin{equation}
	\psi \left( \mathbf{r},t\right) =
	\begin{pmatrix}
		\psi_{1}\left( \mathbf{r},t\right) \\
		\psi_{2}\left( \mathbf{r},t\right) \\
		\psi_{3}\left( \mathbf{r},t\right) \\
		\psi_{4}\left( \mathbf{r},t\right)
	\end{pmatrix}.
	\label{d4a}
\end{equation}
As a result, the Dirac equation (\ref{d1}) with the Hamiltonian (\ref{d2}) is a matrix equation
that is equivalent to a system of four first order partial differential equations
(see also Figure~3 for Fermi's original handwritten notes).

The standard harmonic plane wave solution 
has the form:
\begin{equation}
	\psi
	=
	\psi \left(\mathbf{r}, t\right)
	=
	e^{
		\frac{ \iun }{\hbar}
		(\boldsymbol{p} \cdot \mathbf{r}-Et)
	}u
	=
	\exp \left(
		\frac{ \iun }{\hbar}
		(\boldsymbol{p\cdot} \mathbf{r}-Et)
	\right)
	\begin{pmatrix}
		u_{1} \\
		u_{2} \\
		u_{3} \\
		u_{4}
	\end{pmatrix}
	,
	\label{d7}
\end{equation}
where $u$ is a constant four-component complex vector (bi-spinor)
and $p_1$, $p_2$, and $p_3$ are the components of linear momentum $\boldsymbol{p}$ in the $x$, $y$, and $z$ directions.
We choose the eigenfunctions with definite energy and linear momentum of the commuting energy and momentum operators:
\begin{equation}
	 \iun \hbar \diffp{}{t} \psi = E \psi,
	\qquad
	\boldsymbol{\widehat{p}} \psi = -\iun\hbar \nabla \psi = \boldsymbol{p} \psi .
	\label{d7a}
\end{equation}
Substitution into the Dirac equation results in an eigenvalue problem:
\begin{equation}
	\left(
		c
		\boldsymbol{\alpha} \cdot \boldsymbol{p}
		+mc^{2}
		{\beta}
	\right)
	u
	=
	Eu,
	\label{d8}
\end{equation}
where $E$ is the total energy and $\boldsymbol{p}$ is the linear momentum of the free electron.

One may rewrite the same equation in a block form using the partitioned matrices:
\begin{equation}
	\begin{pmatrix}
		(mc^2-E) I 
        & c \boldsymbol{\sigma} \cdot \boldsymbol{p} \\
		c \boldsymbol{\sigma} \cdot \boldsymbol{p} & -(mc^2+E) I 
	\end{pmatrix}
	u = 0.
	\label{d9}
\end{equation}
Although Fermi introduces the Dirac equation in detail in his notes \cite{Fermi}, lecture 34, he does not elaborate on solving it for a free particle. Therefore, our immediate goal is to demonstrate that this representation is particularly convenient for finding eigenvalues and the corresponding eigenvectors using simple matrix algebra.

The eigenvalue problem (\ref{d8})--(\ref{d9}) can be solved with the help of the following matrix identity:
\begin{eqnarray}
&&\left(
\begin{array}{cc}
\left( mc^{2}-E\right) I 
& c\boldsymbol{\sigma \cdot p} \\
c\boldsymbol{\sigma \cdot p} & -\left( mc^{2}+E\right) I
\end{array}%
\right) \left(
\begin{array}{cc}
\left( mc^{2}+E\right) I 
& c\boldsymbol{\sigma \cdot p} \\
c\boldsymbol{\sigma \cdot p} & -\left( mc^{2}-E\right) I
\end{array}%
\right) \label{ma1} \\
&&\qquad \qquad \qquad \qquad \qquad \qquad \quad \qquad =\left(
m^{2}c^{4}+c^{2}\boldsymbol{p}^{2}-E^{2}\right) \left(
\begin{array}{cc}
I & O \\
O & I
\end{array}%
\right) . \notag
\end{eqnarray}%
Here, the first $4\times 4\/$ partitioned matrix on the left-hand side is identical to that in (\ref{d9}), while the second matrix is derived by substituting $E\rightarrow -E\/.$ The well-known identity, 
$\left( \boldsymbol{\sigma \cdot p}\right) ^{2}=\boldsymbol{p}^{2}%
\ I \/$ \cite{Dirac75},
plays a crucial role in simplifying the matrix multiplication in (\ref{ma1}) and in demonstrating that these matrices commute.
The second observation is the following: a `mixed' partitioned matrix, where
the first two columns of the second matrix are combined with the last two
columns of the first one, has a diagonal square, namely,
\begin{eqnarray}
&&\left(
\begin{array}{cc}
\left( mc^{2}+E\right) I
& c\boldsymbol{\sigma \cdot p} \\
c\boldsymbol{\sigma \cdot p} & -\left( mc^{2}+E\right) I
\end{array}%
\right) \left(
\begin{array}{cc}
\left( mc^{2}+E\right) I
& c\boldsymbol{\sigma \cdot p} \\
c\boldsymbol{\sigma \cdot p} & -\left( mc^{2}+E\right) I
\end{array}%
\right) \label{ma2} \\
&&\qquad \qquad \qquad \qquad \qquad \qquad \qquad =\left( \left(
mc^{2}+E\right) ^{2}+c^{2}\boldsymbol{p}^{2}\right) \left(
\begin{array}{cc}
I & O \\
O & I%
\end{array}%
\right) . \notag
\end{eqnarray}%
These two elementary facts give a standard solution of the above eigenvalue problem.

Indeed, according to the first identity (\ref{ma1}), all four column vectors of the
second matrix automatically give some eigenvectors provided%
\begin{equation}
m^{2}c^{4}+c^{2}\boldsymbol{p}^{2}=E^{2},\qquad \text{or}\qquad E=E_{\pm
}=\pm R,\qquad R=\sqrt{c^{2}\boldsymbol{p}^{2}+m^{2}c^{4}}. \label{ma4}
\end{equation}%
The rank of this matrix is two due to Theorem~4 on p.~47 in \cite%
{GantmacherMat} and only two of those column vectors are linearly independent (see below for a direct verification).
However, our second properly normalized unitary matrix is, as expected, nonsingular.
It provides all four linearly independent column vectors.
Thus, in an explicit matrix form,
\begin{eqnarray}
&&U = \left( u^{\left( 1\right) },u^{\left( 2\right) },u^{\left( 3\right)
},u^{\left( 4\right) }\right) =\sqrt{\frac{mc^{2}+R}{2R}}\left(
\begin{array}{cc}
I
& \dfrac{c\boldsymbol{\sigma \cdot p}}{mc^{2}+R} \\
\dfrac{c\boldsymbol{\sigma \cdot p}}{mc^{2}+R} & - I
\end{array}%
\right) \notag \\
&&\quad =\sqrt{\frac{mc^{2}+R}{2R}}\left(
\begin{array}{cccc}
1 & 0 & \dfrac{cp_{3}}{mc^{2}+R} & \dfrac{c\left( p_{1}- \iun p_{2}\right) }{%
mc^{2}+R} \\
0 & 1 & \dfrac{c\left( p_{1}+ \iun p_{2}\right) }{mc^{2}+R} & \dfrac{-cp_{3}}{%
mc^{2}+R} \\
\dfrac{cp_{3}}{mc^{2}+R} & \dfrac{c\left( p_{1}- \iun p_{2}\right) }{mc^{2}+R} &
-1 & 0 \\
\dfrac{c\left( p_{1}+ \iun p_{2}\right) }{mc^{2}+R} & \dfrac{-cp_{3}}{mc^{2}+R} &
0 & -1%
\end{array}%
\right)\/ . \label{ma6}
\end{eqnarray}%
Here, the first two columns are the eigenvectors corresponding to the positive energy eigenvalues
$E=E_{+}=R=\sqrt{c^{2}\boldsymbol{p}%
^{2}+m^{2}c^{4}}$ (twice) with the projection of the spin on the third axis $%
\pm 1/2,$ in the frame of reference when the particle is at rest, $%
\boldsymbol{p}=\boldsymbol{0},$ respectively; whereas the normalized
eigenvectors in the last two columns correspond to the negative energy eigenvalues $E=E_{-}=-R=-\sqrt{c^{2}%
\boldsymbol{p}^{2}+m^{2}c^{4}}$ (twice), once again with the projection of
the spin on the third axis $\pm 1/2,$ when $\boldsymbol{p}=\boldsymbol{0},$
respectively.
It's worth noting that the matrix manipulation above allows us to bypass a traditional evaluation of the $4\times 4$ determinant in (\ref%
{d9}), resulting into the factorization of the fourth order characteristic polynomial; say by utilizing the familiar formulas of Schur \cite{GantmacherMat}.
(More details can be found in \cite{GorBarSus23}.)

\section{Last Lecture Notes}

Enrico Fermi's scientific mistakes had a {\textquotedblleft monumental character\textquotedblright\/} ---
such as the controversy surrounding his Nobel Prize
\footnote{see the originals:
\cite{FermiNL38, HahnStrass39, MeitnerFrisch39, Frisch39, Andersonetal39} and \cite{FermiCU55, Schwartz} for the details; a modern review on transuranic elements is given in \cite{Kragh18}.}
and the xenon nuclear reactor poisoning \cite{Schwartz} ---
likely stemming from the novelty and originality of his research during that transformative period in the development of nuclear physics.
(Every physicist would be happy to make such a mistake!)
At a {\textquotedblleft personal level\textquotedblright\/} the initial calculation of the {\textquotedblleft double window effect\textquotedblright\/}
in his family Chicago winter house wasn't that successful either \cite{FermiLaura}. But this list is rather short!
On the contrary, Fermi freely shared his deep scientific ideas with colleagues and later was not involved in their publications, some at the Nobel Prize level
\cite{Schwartz}.%
\smallskip
We are examining Fermi's original lecture notes for the course Physics 341/342: Quantum Mechanics, delivered at the University of Chicago \cite{Fermi} during the winter and spring quarters of 1954. Spanning approximately sixty lectures across two quarters, these notes capture the essence of Fermi's teaching. At the end of each lecture, Fermi would compose a problem closely related to the material he had just covered that day (this list of problems is available in the Second Edition). For additional details on some topics, Fermi occasionally referred to Leonard Schiff's {\textit{Quantum Mechanics\/}}, First Edition \cite{Schiff}, as well as Enrico Persico's {\textit{Fundamentals of Quantum Mechanics\/}}, First Edition \cite{Persico}. (Persico, a close friend of Fermi since childhood in Italy \cite{Segre, SegreHoerlin}, was acknowledged in this work.) 
Originally distributed only among Fermi's students as course material, these lecture notes have since been made widely available, with two English editions and an extended Russian translation
{\footnote{with some typos corrected; see, for example, the definition of matrix multiplication on page~14--5,
equation~(24), in the handwritten notes\/}}
\cite{Fermi}, ensuring their accessibility to the broader physics community.
\begin{figure}[h!]
	\includegraphics[width=1.025\linewidth]{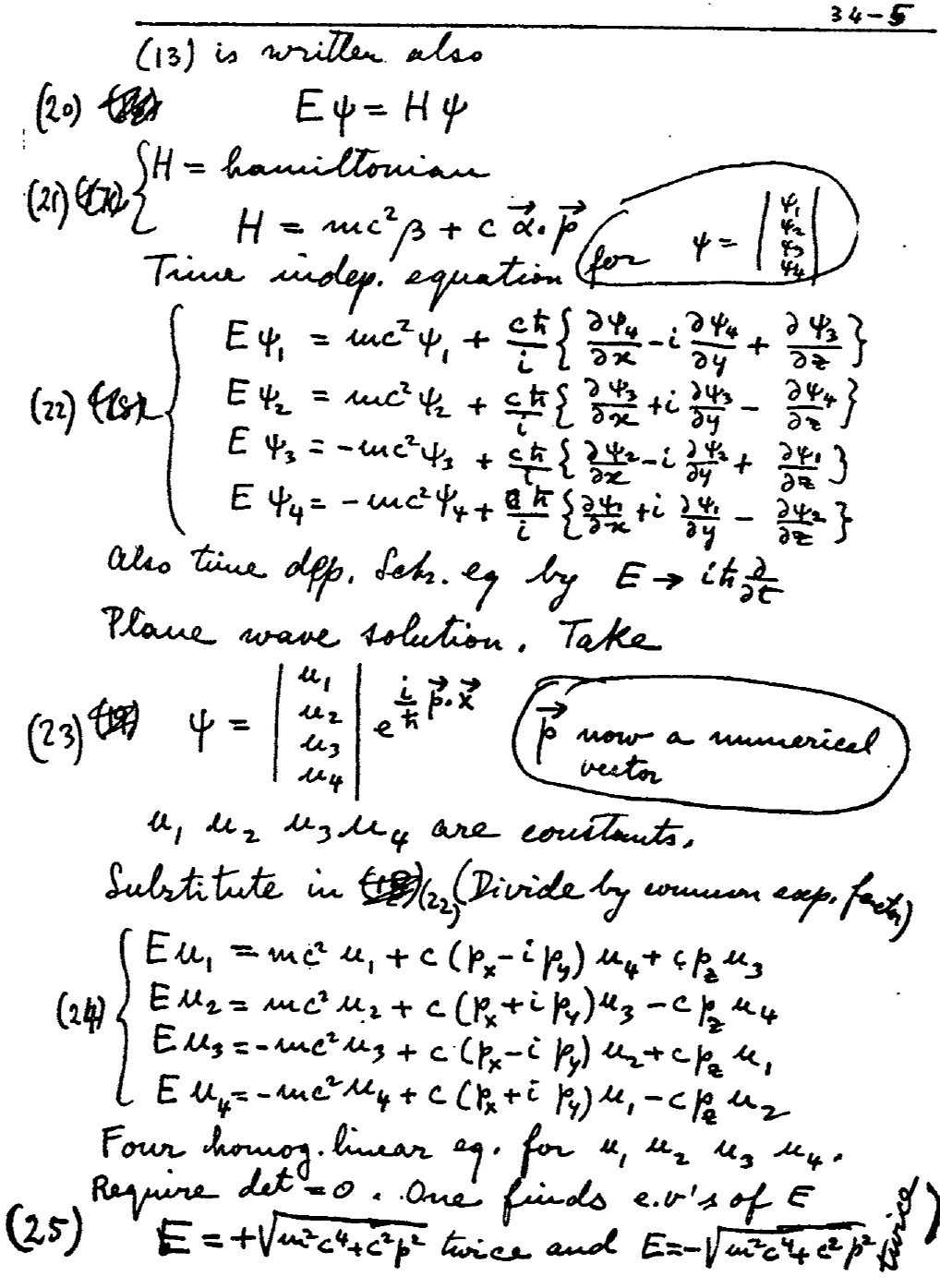}
	\caption{ Page 34--5 of Fermi's lecture notes \cite{Fermi}: Dirac's equation and the eigenvalue problem. }
\end{figure}
%


%
\begin{figure}[h!]
	\includegraphics[width=1.015\linewidth]{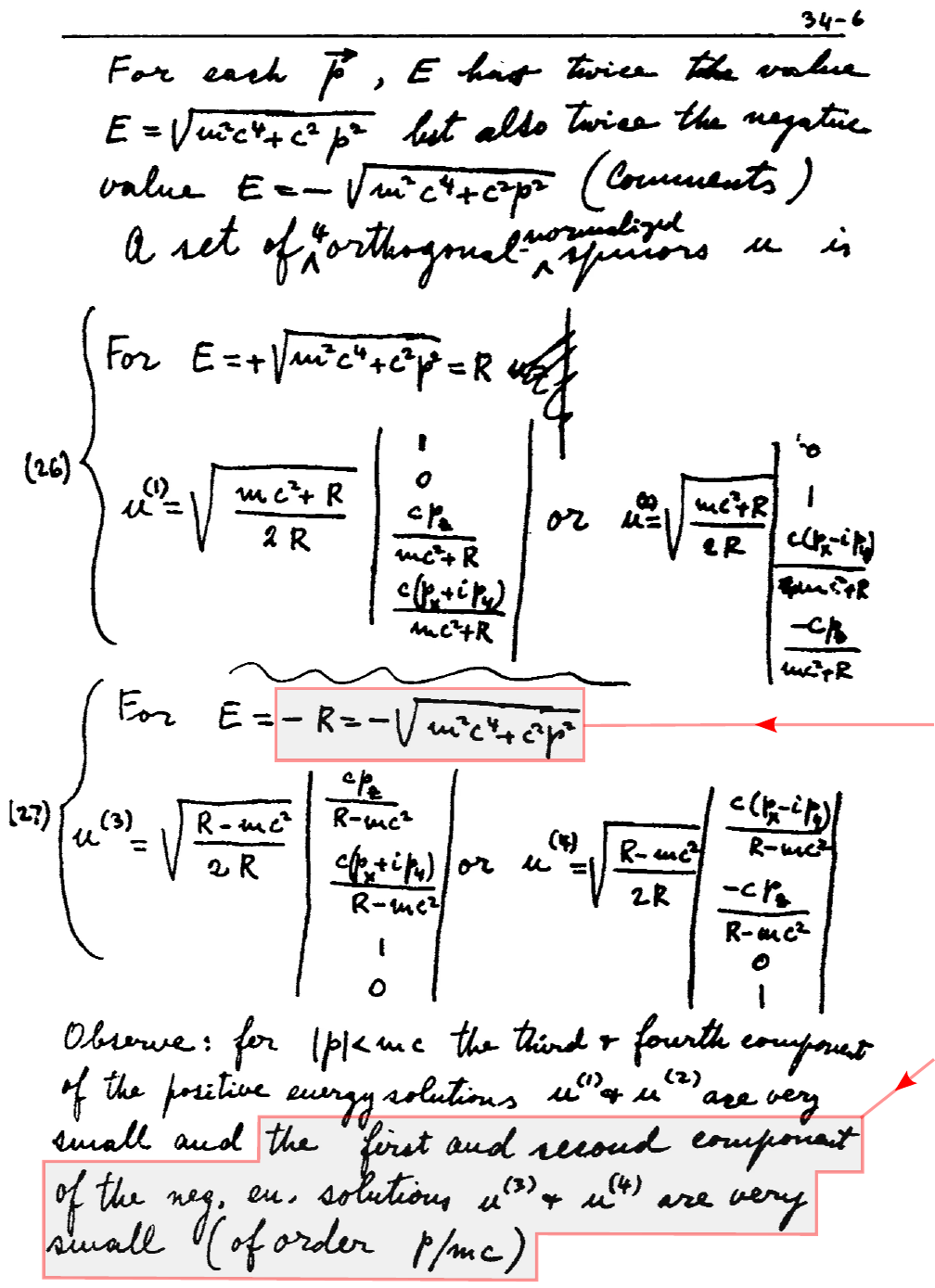}
	\caption{ Page 34--6 of Fermi's lecture notes \cite{Fermi}: observe that both bi-spinors (27) also correspond to $E=+R\/$. }
\end{figure}
\smallskip
We only discuss his lecture~34 on the free relativistic electron.
Our concern is the following: there is an unfortunate mistake, namely,
as it will be shown (see also \cite{GorBarSus23}),
all four bi-spinors (26)--(27) on p.~34-6 for free spin-$1/2$ particle in \cite{Fermi}
(see Figures~3--4 for Fermi's handwritten notes
and Appendix~B in \cite{GorBarSus23}) correspond to the positive energy
eigenvalues $E=E_{+}=+R=\sqrt{c^{2}\boldsymbol{p}^{2}+m^{2}c^{4}}.$
This fact can be verified by a direct substitution into equation (24) on p.~34-5.
For example, in the case of the third bi-spinor $u^{\left( 3\right) }$ given
by equation (27) in Fermi's notes (Figure~4), one gets, up to a constant, that%
\begin{eqnarray}
&&\left(
\begin{array}{cccc}
mc^{2} & 0 & cp_{3} & c\left( p_{1}-\mathrm{i}p_{2}\right) \\
0 & mc^{2} & c\left( p_{1}+\mathrm{i}p_{2}\right) & -cp_{3} \\
cp_{3} & c\left( p_{1}-\mathrm{i}p_{2}\right) & -mc^{2} & 0 \\
c\left( p_{1}+\mathrm{i}p_{2}\right) & -cp_{3} & 0 & -mc^{2}%
\end{array}%
\right) \left(
\begin{array}{c}
\dfrac{cp_{3}}{R-mc^{2}} \\
\dfrac{c\left( p_{1}+\mathrm{i}p_{2}\right) }{R-mc^{2}} \\
1 \\
0%
\end{array}%
\right)  \notag \\
&&\quad =\left(
\begin{array}{c}
cp_{3}\left( \dfrac{mc^{2}}{R-mc^{2}}+1\right) =R\dfrac{cp_{3}}{R-mc^{2}}
\\
c\left( p_{1}+\mathrm{i}p_{2}\right) \left( \dfrac{mc^{2}}{R-mc^{2}}%
+1\right) =R\dfrac{c\left( p_{1}+\mathrm{i}p_{2}\right) }{R-mc^{2}} \\
\dfrac{c^{2}\boldsymbol{p}^{2}}{R-mc^{2}}-mc^{2}=\dfrac{R^{2}-m^{2}c^{4}}{%
R-mc^{2}}-mc^{2}=R \\
c\left( p_{1}+\mathrm{i}p_{2}\right) \dfrac{cp_{3}}{R-mc^{2}}-cp_{3}\dfrac{%
c\left( p_{1}+\mathrm{i}p_{2}\right) }{R-mc^{2}}=0%
\end{array}%
\right) = R\left(
\begin{array}{c}
\dfrac{cp_{3}}{R-mc^{2}} \\
\dfrac{c\left( p_{1}+\mathrm{i}p_{2}\right) }{R-mc^{2}} \\
1 \\
0%
\end{array}%
\right).  \label{cf1}
\end{eqnarray}%
Although Fermi's bi-spinors are normalized, they are not mutually orthogonal.
For example,%
\begin{equation}
\left( u^{(1)}\right) ^{\dag }u^{(3)}=\dfrac{p_{3}}{\left\vert \boldsymbol{p}
\right\vert }\/, \qquad
\left( u^{(2)}\right) ^{\dag }u^{(3)}=\dfrac{ p_{1}+\mathrm{i}p_{2} }{\left\vert \boldsymbol{p}
\right\vert }\/
\end{equation}
(``nobody's perfect"!?). As a result,
\begin{equation}
u^{(3)}=  {\dfrac{p_{3}}{\left\vert \boldsymbol{p}
\right\vert }}\ u^{(1)} +{\dfrac{ p_{1}+\mathrm{i}p_{2} }{\left\vert \boldsymbol{p}
\right\vert }}\ u^{(2)}\/
\end{equation}
%
and
\begin{equation}
u^{(4)}=  {\dfrac{ p_{1}-\mathrm{i}p_{2} }{\left\vert \boldsymbol{p}
\right\vert }}\ u^{(1)} -{\dfrac{p_{3}}{\left\vert \boldsymbol{p}
\right\vert }}\ u^{(2)} \/ ,
\end{equation}
%
as one can easily verify in a similar fashion.
In terms of the helicity operator,%
\begin{equation}
\left( u^{(3)},u^{(4)}\right) =\left( u^{(1)},u^{(2)}\right) \left(
\boldsymbol{\sigma \cdot n}\right) ,\quad \boldsymbol{n=}\dfrac{\boldsymbol{p%
}}{\left\vert \boldsymbol{p}\right\vert }\/ .
\end{equation}

%
The correct answer can be obtained by replacing $R \to -R\/$ in the last two Fermi's original bi-spinors,
with a proper change of the normalization; see (\ref{ma6}).
This result is presented, for instance, in \cite{Persico} and \cite{Schiff} (with $H\to -H$), and verified
in the Mathematica file posted at {\url{https://community.wolfram.com/groups/-/m/t/2933767}}.
%

Once again, Fermi's original bi-spinors are linearly dependent because the
corresponding $4\times 4$ determinant equals zero 
(see also our complementary Mathematica notebook for the detailed calculations).
\smallskip

In his lecture notes (see Figure~4), Fermi also states that: \textquotedblleft for $\lvert
p\rvert \ll mc$, the third and fourth components of the positive energy
solutions $u^{(1)}$ and $u^{(2)}$ are very small; and the first and second
components of the negative energy\footnote{%
As is already noted, all four Fermi's bi-spinors correspond to the
positive energy eigenvalues.} solutions $u^{(3)}$ and $u^{(4)}$ are also
very small (on the order of $p/{mc})$\textquotedblright . This is true only
for the first two bi-spinors. On the contrary, one can easily verify that,
in Fermi's normalization,%
\begin{equation}
\left( u^{\left( 3\right) },\ u^{\left( 4\right) }\right) \rightarrow \left(
\begin{array}{c}
\dfrac{\boldsymbol{\sigma \cdot p}}{\left\vert \boldsymbol{p}\right\vert }
\\
{O}%
\end{array}%
\right) =\left(
\begin{array}{cc}
\dfrac{p_{3}}{\left\vert \boldsymbol{p}\right\vert } & \dfrac{p_{1}-ip_{2}}{%
\left\vert \boldsymbol{p}\right\vert } \\
\dfrac{p_{1}+ip_{2}}{\left\vert \boldsymbol{p}\right\vert } & -\dfrac{p_{3}}{%
\left\vert \boldsymbol{p}\right\vert } \\
0 & 0 \\
0 & 0%
\end{array}%
\right) ,\quad c\rightarrow \infty . 
\end{equation}%

%
It remains unclear how these mistakes originated with Fermi himself and why neither his students nor his publishers corrected them. Interestingly, there is a similarity in notation with Persico's book \cite{Persico}, which may have influenced Fermi's work. 
Notably, equations (113)--(115a,b) in his original work \cite{FermiRad} are equivalent, in matrix form, to our identity (\ref{ma1}); see the complementary Mathematica notebook for further details.
\smallskip

From our `matrix perspective', Fermi took the four properly normalized columns of the second matrix in (\ref{ma1}) as
his eigenvectors. But the rank of this matrix is two. Indeed, let us
subtract from the second row of blocks of this matrix the first one, multiplied
on the left by $(c\mathbf{\sigma \cdot p})/(mc^{2}+E)$ (the generalized
Gaussian elimination algorithm \cite{GantmacherMat}).
As a result,%
\begin{eqnarray}
\left(
\begin{array}{cc}
(mc^{2}+E)I & c\mathbf{\sigma \cdot p} \\
c\mathbf{\sigma \cdot p} & -(mc^{2}-E)I%
\end{array}%
\right)  &\rightarrow &\left(
\begin{array}{cc}
(mc^{2}+E)I & c\mathbf{\sigma \cdot p} \\
O & (E-mc^{2})I-\dfrac{(c\mathbf{\sigma \cdot p})^{2}}{mc^{2}+E}%
\end{array}%
\right)   \notag \\
&\rightarrow &\left(
\begin{array}{cc}
(mc^{2}+E)I & c\mathbf{\sigma \cdot p} \\
O & O%
\end{array}%
\right) .
\end{eqnarray}%
The row \'{e}chelon form explicitly demonstrates that there are only two linearly independent rows and only two linearly independent columns.
\smallskip

It appears that a more general concept of the helicity states for electrons and positrons and the corresponding polarization density matrices
were never discussed in Fermi's introductory quantum mechanics course.
The bottom line is that those formulas, unfortunately, were copied without proper verification into the second English edition
of the lecture notes and also appeared in the Russian translation, which has extensive comments on Fermi's original handwritings.
Some physicists believe that the second formula, for $\alpha\/,$ in a well-known photograph from the Encyclopedia of Britannica (Figure~1),
is, most likely, Fermi's idea of a joke (or he was just having a bad day?) \cite{Huber, Lien, Schwartz}. 
But it does not seem to be the case here?!
Dick Askey used to like to say that, ``if there is a formula in a book (or lecture notes --- SKS), something like that should be true",
meaning that one has to derive this result on her or his own.
\smallskip

In the last three lectures~35--36--37, Fermi discussed introduction of the electromagnetic field into the theory of relativistic Dirac particles,
motion in central field and the ground state of hydrogen atom, as well as the nonrelativistic limit, magnetic moment, and Thomas correction.
Lorentz invariance and charge conjugation are also studied, although
these topics are usually discussed in the quantum field theory courses \cite{AkhBer, BerLifPit, Moskalev, PeskSchroe}.
\smallskip

Relativistic helicity states of Dirac's free particles and the polarization density matrices are discussed in detail in
\cite{AkhBer, BerLifPit, GorBarSus23, Moskalev, PeskSchroe}.
Relativistic Coulomb problems are considered in \cite{AkhBer, BerLifPit, Dav, EllisKoutschanSuslov23, Fermi, Ni:Uv, Schiff, Sus24, Susetal20}
(see also the references therein).
Computer algebra methods are utilized in \cite{EllisKoutschanSuslov23, GorBarSus23}, and elsewhere
(see also our complementary Mathematica notebook for the detailed evaluation of all formulas discussed in this and previous sections and more).
This approach is motivated by an introductory course in mathematics of quantum mechanics which the
author has been teaching at Arizona State University for more than two
decades (see also \cite{Barleyetal2021, EllisKoutschanSuslov23, GorBarSus23, Sus24, Susetal20}
for more information).
\smallskip

\noindent \textbf{Acknowledgments.\/} The author is grateful to Ruben Abagyan, Maria Aksenteva, Kamal Barley, Victor Dodonov, Ben Goren, Sergey Kryuchkov, Nathan Lanfear, Oleg Poluektov, Cameroon Reed, Andreas Ruffing, Eugene Stepanov, Erwin Suazo, Jose Vega-Guzm\'{a}n, and Alexei Zhedanov for valuable comments, discussions, and help.

\bigskip
{\small{{{\scshape{Orcid:\/}}
{{\url{https://orcid.org/0000-0001-8169-0987}}}}}}

\end{document}